# Density of States Proportion on Charge Transfer Kinetics in Breathing Fermionic Systems of Molecules and Materials: A Perspective of Entropy-Ruled Method

K. Navamani*

Department of Physics, Centre for Research and Development (CFRD)

KPR Institute of Engineering and Technology

Coimbatore-641 407, India.

*Corresponding author: pranavam5q@gmail.com

**Abstract**

Conceptualization, theory/method developments and implementations are always of great importance and an interesting task to explore a new dimension in science and technology, which is highly solicited for various functional-driven potential applications (e.g., electronic devices, charge storage devices). Numerous experimental and theoretical studies urge the necessity of a new theory or method to quantify the exact value of charge transport (CT) calculations (e.g., mobility and conductivity) through the appropriate process and methods. With this motivation, the entropy-ruled charge dynamics method has been recently proposed, which unifies band and hopping transport mechanism via quantum-classical transition analogy. Here, the energy (in terms of chemical potential) scaled entropy has a direct proportion with the density of states (DOS); and hence it is termed as *DOS proportion*. This proportion principally acts as a key descriptor for charge transport calculations in both molecular and materials systems, which is directly connected with all CT quantities like mobility, conductivity, current density etc. This perspective explains a unique nature of entropy-ruled method for the entire transport range from delocalized band to localization (or hopping) transport. The validity and limitations of Einstein's relation and Boltzmann approach are discussed with different limits and physical conditions for disordered molecules and periodic systems. Finally, the futuristic scope and expected progress is addressed for correlated electron dynamical systems and devices. It is well-noted that the new DOS proportion and related entropy-ruled transport formalism are fundamentally more important for nurturing semiconducting science and technology towards a new era.

## 1. Introduction

Charge transport (CT) is a fundamentally important mechanism for all typical systems from molecules to materials to explore and categorize the various potential applications such as electronic, optoelectronic, spintronic, charge and energy storage devices, thermoelectric and other semiconducting devices.[1,2] Here, the structure-property relation is one among the key factors to describe the electron and energy distributions in any system of interest, which is principally associated with the device functionality.[3-5] Especially, the size, shape, nature of the chemical compound, structural order or disorder, and other geometric property are mainly controlling the charge and energy transport behaviour, which directly involves designing the suitable functional devices. In this regard, the microscopic understanding of charge dynamics (phase space effect) at a given system (molecules/materials) is highly essential and solicited. The intrinsic picture of the systems or devices deals with the electronic (energy) levels distribution in a phase space, and is principally correlated with molecular or chemical structure/compound and geometric pattern.[5-8] This intrinsic picture of any material can be analysed by electronic structure calculation via quantum mechanical principles or relevant conventional methods.[5,6] The electronic structure study normally reveals the delocalized band information and density of states, which are fundamentally connected with the electronic transport.

The external influences (by electric, magnetic, doping process and other environmental and thermodynamic effects, including disorder nature) along with the intrinsic quantum picture (i.e., electronic structure) usually decides the nature of electron hole dynamics/transport and its related energy flux within a particular molecular or material systems.[9-11] The extrinsic (e.g., bias due to electric or magnetic field) effect on intrinsic or electronic structure of the system elucidates drift coupled charge density flux and related energy transport.[4,12,13] Here, the density flux facilitates the diffusion along the consequential electronic sites of the systems. In this context, the understanding of structure-property relationship is crucial for any potential applications. Here, the intrinsic picture of the molecular/material systems (i.e., electronic structure)-CT property relation is a highly solicited study for electronic, photonic and energy devices applications.[1,8] The electronic level analysis is closely connected with the localization or delocalization of the charge and density of states of the given systems.[2,14] For particular applications, one can attain the relevant functional behaviour through electronic structure-property relationship with the help of external features like bias (e.g., electric or magnetic field), temperature, doping etc. In this connection, there is a two main classification of systems such as ordered (periodic) and disordered systems; accordingly, the electron/hole dynamic mechanism changed. In principle, the charge dynamics in periodic materials follows band transport (adiabatic delocalization) theory, and disordered systems underlie the hopping transport mechanism (nonadiabatic localization). It is generally anticipated that the coincides of many local minima in the potential energy surface of ordered materials, which causes the electronic compressible delocalized energy band.[8,15] On the other hand, incompressible electronic states are observed in the disordered solids due to misaligned energy landscapes. Here, the electronic compressibility is a direct consequence of the density of states (DOS).[16,17]

Numerous experimental and theoretical reports further indicate the possibility of crossover transport between band and hopping mechanism for various systems, which mainly depends on the amount of disorder (weak to strong) on charge transport.[18-23] Especially, the interaction between electronic and nuclear or lattice dynamics gives a significant impact on CT principles via static to dynamic disorder transition. Relatively fast nuclear dynamics (faster than electron dynamics) is responsible for dynamic disordered transport; it led towards bank-like transport. In such dynamic disordered transport cases, the expected oscillations frequencies (fluctuations) are in the order of harmonic frequencies, in which the large polaron band theory is suitable to determine the electronic transport.[11,23] The small polaron hopping transport mechanism is more fit for static disordered systems. However, the time scale of lattice vibration (nuclear dynamics), and typical interactions between electronic and nuclear degrees of freedom are accountable for band-to-hopping transport transition (crossover relation).[8,17,23,24] In this regard, over the past few decades, some of important transport mechanism/models such as surface hopping model,[20,25] flickering resonance method,[26,27] transient localization transport,[28] diffusion limited by dynamic disorder,[7] density flux model for hopping conductivity,[29] and forth-back oscillated diffusion[30] etc., were proposed by many scientists to aim of exploration for the exact CT calculation. The above models and methods have their own merits; although we need a unique and more-accurate approach to cover the entire range of CT mechanism from band to hopping (via intermediate) at wide physical conditions.

Especially for field effect transistor (FET) applications, the challenging task is to predict the cooperative nature between electronic degeneracy and thermodynamics on mobility.[31,32] Besides that, it is expected for periodic band materials, the lattice vibration is quantized (localized) known as phonon, and charge has a delocalized behaviour. On the other side, the disordered systems (e.g., molecules) are possibly having dispersed nature (delocalised) of lattice or nuclear dynamics due to non-periodic oscillation/dynamics, and hence the charge carrier is localized. This inverse characteristic proportion between band and hopping transport systems further makes it interesting, but it is a challenging task to incorporate both of them. Hence, the relaxation time and hopping time (i.e., inverse of charge transfer rate) are inverse relation with each other, and these are the fundamental and direct proportional entities of band and hopping mobility, respectively.[13,33] However, the lattice/nuclear dynamics coupled electron transport possibly leads to the intermediate (or crossover) transport between hopping and band mechanism, which can be modified by cooperative behaviour of temperature, doping or substitution, applied bias and structural fluctuation. In such that for any CT systems (ordered or disordered systems), the electronic structure and thermodynamics effects are playing as key descriptors for mobility and other CT parameters calculations, which can be principally accounted for by quantum-classical dynamics pictures.[10,32,34]

To overcome the above drawbacks and to explore some unknown key functional factors in CT mechanism, the scientific community primarily requires a new and unique method/approach to take-up the next generation of advanced electronic and semiconductor technology. Keeping with these points along with the knowledge of numerous research reports/studies over a half century, the entropy-ruled CT method has been systematically developed with the concern of various factors like equilibrium, nonequilibrium, bias-driven degeneracy weightage, disorder and correlated electron behaviour, etc.[13,17,22,32,35,36] This is a unique method for both quantum and classical systems to investigate the exact CT calculation in the entire regime from band-to-hopping. In this perspective, the first two sections cover the description of nonequilibrium fluctuation theory-based entropy production rule and its consequence effect on charge density for different dimensional (1D, 2D and 3D) systems. Thereafter, the DOS proportion is introduced via electronic compressibility relation which is directly proportionate with DOS, it also describes bias-driven degeneracy effects on DOS at various temperature situations. The DOS proportion is defined as the changes of effective entropy ($h_{eff}$) with respect to the chemical potential ($\eta$). The simplified definition of DOS proportion is described as the entropy variation per unit energy of the system. Here, the effective entropy is generally described as the combinatory effect of all typical entropy, which mainly includes the differential entropy and thermodynamic entropy. The charge distribution and dynamics can be studied for molecular and material systems using DOS proportion, which is directly associated with the electronic and charge storage devices' functionalities. To this extent, the fundamental relationship between DOS proportion and diffusion-mobility relation (D/μ) is explained; accordingly, developed other transport quantities like conductivity, current density, diode ideality factor are illustrated in this perspective. To this extent, the validity of entropy-ruled method for band and hopping transport systems are shown and verified in equilibrium as well as in nonequilibrium cases using four-sets of analytical procedures through some molecular solids and material systems. Finally, the future scope and expected developments with few implications are discussed and addressed.

## 2. Nonequilibrium Carrier Energy Flux-Based Entropy Production Rule and Charge Density Variation

As we know that the total energy of any electronic system is the summation of kinetic, and other interactions potential energy (including exchange and correlations). During CT, the energy flux also occurred along with the carrier motion, which can be studied by momentum and energy redistribution. This carrier energy flux possibly takes from equilibrium to nonequilibrium or quasi-equilibrium transport (i.e., nearer to equilibrium), which depends on magnitude of traversing carrier energy (or chemical potential) along the consequential electronic levels. Here, the chemical potential shift leads to charge density flux and it facilitates the diffusion-based transport mechanism; since the chemical potential usually has a direct relationship with carrier concentration. In this regard, the change in energy over the time interval of an electronic system during CT is fundamentally important to explore the electronic property. That is, the amount of energy changes per unit time (or energy flux rate) can provide a more accurate CT estimation at a given system. This energy flux with time has direct association with the drift coupled diffusion transport, which is mostly expected in degenerate systems and real devices. With this motivation, the derived expression for the variation in rate of change of energy flux at different physical conditions is described by,[13,35,37]

$$\frac{\Delta E}{\Delta t}\bigg|_{h_{rel}} = \frac{\Delta E}{\Delta t} exp\left(-\frac{S_{rel}}{k_B}\right) \qquad (1)$$

where, $\frac{\Delta E}{\Delta t}\bigg|_{S_{rel}}$ and $\frac{\Delta E}{\Delta t}$ are the final and initial energy flux rate (or rate of change of energy) during CT at two different physical or thermodynamic situations, which is here exponentially compensated by relative entropy ($S_{rel}=S_f-S_i$). For instance, the above energy flux rate equation during CT with and without applied bias (e.g., electric field or different doping concentrations) at fixed temperature ($T$) can be written as $\frac{\Delta E(\vec{E})}{\Delta t}\bigg|_{S_{rel}} = \frac{\Delta E(\vec{E}=0)}{\Delta t} exp(S_{\vec{E}} - S_{\vec{E}=0})$. Similarly, the rate of energy flux equation in two different $T$ values at fixed bias conditions (i.e., constant electric or magnetic field and fixed doping concentration), is expressed as $\frac{\Delta E(T_2)}{\Delta t}\bigg|_{S_{rel}} =$ $\frac{\Delta E(T_1)}{\Delta t} exp(S_{T_2} - S_{T_1})$. Here, the bias voltage-driven chemical potential decreases the thermodynamic entropy, but it increases with the temperature. The charge transfer kinetics is generally drifted by electric field and appropriate doping, which is connected with the chemical potential. On the other hand, the temperature here reduces the carrier energy flux, which can be accounted for by thermodynamic entropy. In this regard, the understanding of CT property through electronic and thermal aspects is essential to functionalize both the charge transport and charge storage devices. The previous reports clearly emphasize that the total effects from the electronic (including degeneracy effect by external bias) and thermal (i.e., temperature) contributions through effective entropy on charge and energy transport.[32,35] Here, the effective entropy is $h_{eff} = h_S - S/k_B$; where, $h_S$ and $S$ are the differential entropy and thermodynamic entropy, respectively. The differential entropy is generally derived from the Gaussian function, $h_S(E) = -\int_{-\infty}^{+\infty} \Phi(\varepsilon) \ln \Phi(\varepsilon) d\varepsilon = \ln(\sigma_{GW}\sqrt{2\pi e})$, where, $\varepsilon$ is the normalized energy, $\varepsilon'/k_B T$ and $\Phi(\varepsilon)$ is the Gaussian function and this form is $\frac{1}{\sigma_{GW}\sqrt{2\pi}} exp\left(-\frac{\varepsilon^2}{2\sigma_{GW}^2}\right)$. In this relation, the differential entropy ($h_S$) is directly associated with the Gaussian width ($\sigma_{GW}$), which describes the charge delocalization property in the systems. That is, the $h_S$ closely connected with the electronic level population. That is, the number of states is normally accounted for by the statistical entropy, $\ln(N_{St})$, and its consequences on transport can be examined by differential entropy $h_S = \ln\left(\sqrt{2\pi e}\sigma_{GW}\right)$. At the same time, the thermodynamic entropy ($S$) gives the information about the thermodynamic disorder and is weighted by the amplitude of thermal vibration (e.g., thermal polaron scattering). The scattering potential by temperature usually limits the carrier transport and hence is expected $\sigma_{GW}$ to be small enough. In effective entropy ($h_{eff} = h_S - \frac{S}{k_B}$), the $h_S$ favors the delocalized CT (continuum band-like) and it also quantifies the field-response effect on CT (i.e., drift mobility).[35] On the other hand, the thermodynamic entropy ($S$) suppresses the electronic coupling and increases the vibronic coupling (due to thermal vibration) which in turn towards the potentially trapping charge localization. Based on the electronic (quantum picture) and thermal (classical) contributions on CT, the generalized form of carrier energy flux rate equation can be expressed as,[13,35]

$$\frac{\Delta E}{\Delta t}_{h_{eff}} = \frac{\Delta E}{\Delta t} exp(h_{eff}) \qquad (2)$$

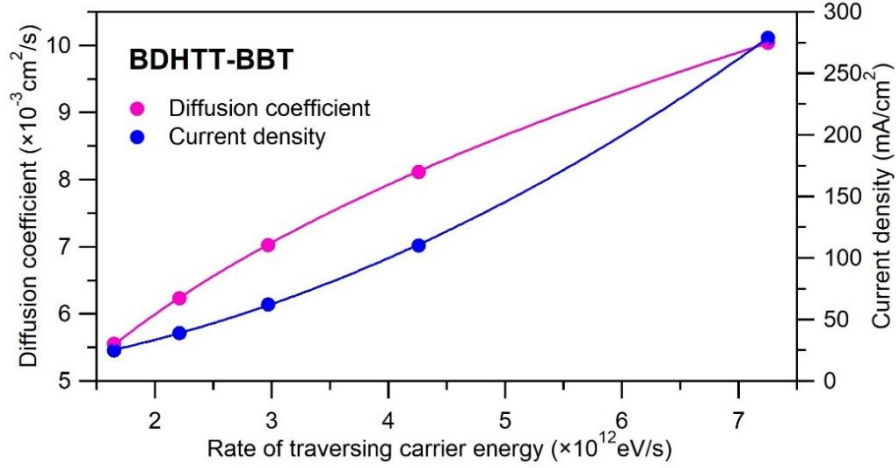

**Figure 1:** Changes in diffusion coefficient (*D*) and current density (*J*) with respect to the rate of traversing carrier energy for BDHTT-BBT molecules at different electric field-assisted site energy differences of 0, 20, 40, 60, and 80 meV. The plot shows that the current density (*J*) is an abruptly varying parameter rather than *D*. The detailed analysis was performed in ref. 13.

For instance, the carrier energy flux rate along the stacked dialkyl-substituted thienothiophene-capped benzobisthiazole (BDHTT-BBT) molecular units directly influences the diffusion and current density which is shown in Fig. 1. In principle, from the above equation (2), one can describe the entropy production even at nonequilibrium fluctuation. Therefore, one another form of bias-driven carrier's energy nonequilibrium fluctuation theory-based entropy production rule as,

$$h_{eff} = ln\left[\frac{\left(\frac{\Delta E(\vec{E},T)}{\Delta t}\right)}{\left(\frac{\Delta E(\vec{E}=0,T)}{\Delta t}\right)}\right] \qquad (3)$$

This relation provides a logarithmic of the ratio between carrier's energy rate at with and without bias (e.g., electric field or doping) conditions. That is, based on the variation in carrier energy with time along the charge percolation network, we can find the entropy contribution for charge dynamics at a given electronic system. By applying the uncertainty relation, the above relation becomes,

$$h_{eff} \cong 2ln\left(\frac{\Delta E(\vec{E},T)}{\Delta E(\vec{E}=0,T)}\right) \qquad (4)$$

Here, the production of entropy can be measured by energy flux variation under two different physical conditions for the same system. From this relation, we can also define the temperature contribution to the entropy production of a given electronic system as the change in total electronic energy per unit thermal energy ($k_B T$). Similarly other physical parameters' (e.g., electric or magnetic, doping, etc.) contribution to the entropy on any Fermionic systems can be quantified. This approach gives a new perspective on electronic transport studies and other relevant property exploration (including phase transition) with respect to the entropically varying thermodynamic effect.

Moreover, one among the key parameters for the carrier mobility calculation in both ordered and disordered systems is deformation potential (for periodic materials) and relaxation or reorganization energy (for disordered molecules). These parameters generally describe geometric relaxation, which generally act as a barrier for charge transport phenomena.[1,2,6] In this regard, the required minimum energy for structural breathing (geometric relaxation), during CT, is crucial to describe the exact structure-property (here, electronic transport property) relationship. The external bias-response in an electronic systems or devices and its consequences on transport property relation can be shown by the structural relaxation over the time interval, which is direct association with equation (2). Total energy of free electron solid systems of bulk (3D), layered (2D) and nanorods or quantum wires (1D) is expressed as,[2,35,38]

$$E_{3D} = \frac{\hbar^2 n_{3D}^{\frac{5}{3}} V}{10\pi^2 m} \rightarrow \frac{\hbar^2 k^5 L^3}{10\pi^2 m} \qquad (5.a)$$

$$E_{2D} \propto \frac{\hbar^2}{10\pi^2 m} n_{2D}^{\frac{5}{2}} A^{\frac{3}{2}} = \frac{\hbar^2 N^{\frac{5}{2}}}{10\pi^2 m} A^{-1} \qquad (5.b)$$

$$E_{1D} \propto \frac{\hbar^2 N^5}{10\pi^2 m} \frac{1}{L^2} \qquad (5.c)$$

Since, the wave vector for 3D, 2D and 1D systems is as follows, $k = \sqrt[3]{3\pi^2 n_{3D}}$, $k = \sqrt{2\pi n_{2D}}$, and $k = \frac{\pi n_{1D}}{2}$, respectively. Here, $\hbar$, $n_{dD}$, $V$, $A$, $L$, $k$ and $m$ are the reduced Planck constant, carrier density of different dimensional systems (d = 1D, 2D and 3D), volume, area, length of an electron systems, wave vector and effective mass of an electron, respectively. The required minimum energy under structural relaxation over the time interval can be estimated by Eqn. 2, along with the energy expressions 5(a)-5(c). Using this approach (inserting Eqn. 5 into Eqn. 2), the governing equation of entropically varying charge density for different dimensional systems is generalized as,[35]

$$n_{h,S} = n_i exp\left[\frac{d}{d+2}\left(h - \frac{S}{k_B}\right)\right] \qquad (6)$$

where, $n_{h,S}$, $n_i$, $d$, $h$, $S$ and $k_B$ are the final charge density which includes entropy effect, initial charge density, dimension (d = 1D, 2D and 3D), differential entropy, thermodynamic entropy and Boltzmann constant, respectively. Interesting observation from this relation is that the dimension dependent-fractional factor, $\frac{d}{d+2}$, is equivalent to that of ratio between average energy ($E_{av}$) and Fermi energy ($E_F$) of the systems, normally varies with the dimension (1/3 for 1D, 1/2 for 2D and 3/5 for 3D).[2,35,39]

In general, charge density has a direct relation with the chemical potential. Here, the energy flux (in terms of chemical potential) of an electronic system causes the carrier density flux, which is a measurable quantity through an effective entropy. This entropy accounts for both the electronic and temperature effects on charge density variation. Besides that, the presence of disorder is responsible for energy variation (inhomogeneous chemical potential) due to nonuniform interaction between the local minima of potential energy surface, which is also accounted for by effective entropy. According to the net scattering and other interactions potential, the energy dispersion ($E$ vs $k$) variation is expected in the Fermionic systems, which is inversely proportional with the entropy, $S(T,k) = \frac{1}{3}Vk_B^2T\left[k^2\left(\frac{\partial E_k}{\partial k}\right)^{-1}\right]_{E_k=\eta}$ (see ref. 22). Moreover, the bias-driven degeneracy effect due to applied electric or magnetic field and doping on electronic behaviour can be included through differential entropy, $h_S = \ln\left(\sqrt{2\pi e}\sigma_{GW}\right)$, which is responsible for quantum flux. Here, the Gaussian width of an electron wave packet decides the electron dynamics, either it follows localization or delocalization at particular Fermionic systems. The effective entropy ($h_{eff} = h_S - \frac{S}{k_B}$) provides a cooperative effect from the thermal and quantum contribution to charge dynamics at wide classes of molecular and material systems. Therefore, one-to-one variation between energy flux (or chemical potential) and entropy essentially required to explore the electronic property of any materials or molecules. Importantly, the effect of change in energy per unit entropy on mobility, density of states, conductivity and current density are foremost important in this method, which is a unique approach for nonadiabatic (localized) and adiabatic band transport mechanism, including intermediate transport systems at entire thermodynamic range.

## 3. Density of States Proportion and Charge Transport Quantities

Charge distribution along the electronic levels is a most fundamental factor to describe electron transfer kinetics in a system. The CT quantities such as mobility and conductivity, etc., are direct incorporation with the nature of energy level population, whether the system has compressible electronic band or incompressible separated states. This number of energy states is usually studied through density of states (DOS). In general, the electronic compressibility is another equivalent form of DOS, which is defined as the change in carrier density (or distribution over the electronic levels) with respect to the chemical potential, $\frac{dn}{d\eta}$. Using generalized form of entropically varying charge density equation, the formulated DOS for 1D, 2D and 3D systems as given by,[35]

$$\left(\frac{dn_{h,S}}{d\eta}\right)_d = \left(\frac{d}{d+2}\right)\left(\frac{dh}{d\eta} - \frac{1}{k_B}\frac{dS}{d\eta}\right) n_i \exp\left[\frac{d}{d+2}\left(h - \frac{S}{k_B}\right)\right] \quad (7.a)$$

Or

$$\left(\frac{dn_{eff}}{d\eta}\right)_d = \left(\frac{d}{d+2}\right)\left(\frac{dh_{eff}}{d\eta}\right) n_{h_{eff}} \quad (7.b)$$

where, $n_{eff}$ is the effective charge density, which has a generalized form of Eqn. 6. It is observed that the DOS is directly proportional to the $\frac{dh_{eff}}{d\eta}$, and hence it is termed as DOS proportion.[40] Here, the DOS proportion is defined as the change in effective entropy per unit chemical potential. According to the new notion, the carrier density times DOS proportion gives the DOS of the system (see ref. 35). The above equation is a new version of DOS, which can also be described as energy (in terms of chemical potential) scaled entropy parameter.

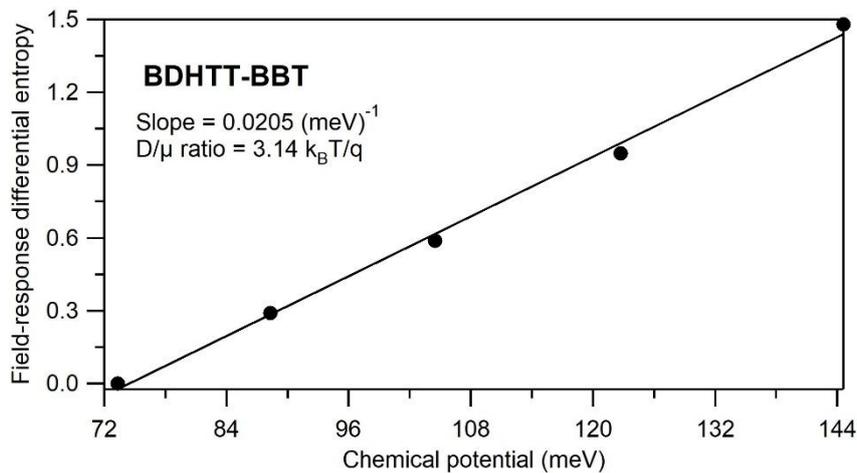

**Figure 2:** The linear enhancement of differential entropy with respect to the chemical potential is observed for the BDHTT-BBT molecule at different site energy fluctuation values of 0, 20, 40, 60, and 80 meV. Closed circles are theoretically computed chemical potential at different site energy flux-associated differential entropy, and the solid line is a fitted plot. The slope value gives a DOS proportion, which is the key descriptor for mobility and conductivity. For more details, see refs. 13 and 33.

This new entity of DOS (i.e., DOS proportion) principally rules charge dynamics in the electronic systems, which can be characterized by mobility, conductivity, current density and diode ideality factor, etc. The DOS proportion was calculated for BDHTT-BTT molecule through chemical potentially driven differential entropy at fixed room temperature, which is plotted in Fig.2. In this connection, the diffusion-based mobility recently proposed using generalized Einstein relation $\left(\mu = q\left(\frac{dn}{d\eta}\right)\frac{1}{n}D\right)$ along with the DOS proportion, and its explicit form as,[35]

$$\mu_d = \left[\left(\frac{d}{d+2}\right)q\frac{dh_{eff}}{d\eta}\right]D \qquad (8)$$

where, $q$ and $D$ are the electric charge and diffusion coefficient, respectively. The mobility equation states that the entropy production per unit chemical potential is a ruling factor for diffusion-mobility relation ($\mu/D$). Therefore, this new method is the so-called entropy-ruled method, which is an alternative version of classical Einstein's relation $\left(\mu_{Einstein} = \frac{q}{k_BT}D\right)$ and suitable for both quantum and classical systems with wide thermodynamic range. It is noted that Einstein's relation ($\mu/D$) does not depend on dimension and linearly varying with the temperature or thermal energy. On the other hand, the entropy ruled transport relation depends on the dimension associated with the fractional value of $d/(d+2)$, along with the DOS proportion. At appropriate conditions (nondegenerate or weakly degenerate ($k_BT \gg \eta$), equilibrium and high temperature), the entropy-ruled relation turns out to be the Einstein relation. In such cases, the DOS is an equivalent to that of $\frac{d+2}{d}\frac{1}{k_BT}$ $\left(i.e., \frac{\Delta h_{eff}}{\Delta \eta} \approx \frac{d+2}{d}\frac{1}{k_BT}\right)$. With respect to the DOS proportion, the $\mu/D$ relation takes either linear or nonlinear behaviour. Here, energy (in terms of chemical potential) scaled effective entropy $\left(i.e., \frac{dh_{eff}}{d\eta} \to DOS\right)$ acts as a ruling descriptor for mobility calculation at given electronic systems or devices (see Fig. 2). In this extent, according to this entropy-ruled method, the governed conductivity equation is

$$\sigma_d(h_{eff}, \eta, D) = Dn_i\left[\left(\frac{d}{d+2}\right)q^2\frac{dh_{eff}}{d\eta}\right]exp\left[\left(\frac{d}{d+2}\right)h_{eff}\right] \equiv q^2Dn_{h_{eff}}\left[\left(\frac{d}{d+2}\right)\frac{dh_{eff}}{d\eta}\right] \qquad (9)$$

In general, the field-response (bias-driven) behaviour in the charge transport devices, like field effect transistors (FET), is analysed by current density ($J$)-voltage ($V$) characteristic study. Here, the device performance and typical transport mechanism (either Langevin or Shockley-Read-Hall) are illustrated through the ideality factor ($N_{id}$), which is incorporated in the diode equation. The diffusion-mobility relation is a main influencing factor in the Shockley diode current density equation, which has an original form as,[12]

$$J = J_0 \left[ exp\left(\frac{qV}{k_BT}\right) - 1 \right] \quad (10.a)$$

From the above diode equation, the presence of $\frac{q}{k_BT}$ comes from classical Einstein's relation, μ/D. The empirical form of diode is given by[12]

$$J = J_0 \left[ exp\left(\frac{qV}{N_{id}k_BT}\right) - 1 \right] \quad (10.b)$$

where, $J_0$ represents the saturation current density. In this extent, using entropy-ruled method, the generalized Shockley diode equation can be written as,[32,35,36]

$$J = J_0 \left[ exp\left(qV\left(\frac{d}{d+2}\right)\frac{\Delta h_{eff}}{N_{id}\Delta \eta}\right) - 1 \right] \quad (11)$$

This diode equation is valid for both nondegenerate and degenerate systems of 1D, 2D, and 3D in the entire transport region from quantum-to-classical at wide temperature range. The fluctuation-associated nonequilibrium diffusion current can be accounted for here by $\Delta h_{eff}$ and $\Delta \eta$. By taking numerical differentiation of this modified Shockley diode equation $\left[N_{id} \cong \left(\frac{d}{d+2}\right)q\frac{dh_{eff}}{d\eta}\frac{dV}{d(lnJ)}\right]$, one can get the diode ideality factor, which is a key factor to evaluate the device performance.[36]

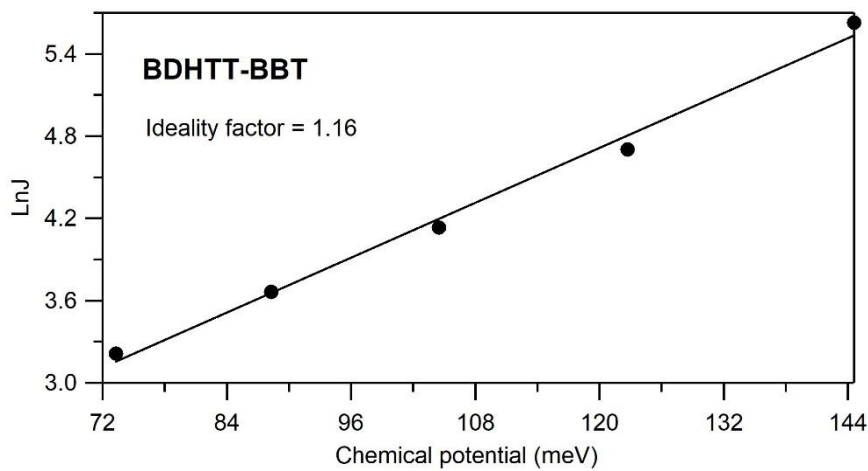

**Figure 3:** Calculated ideality factor (using Eqn. 11) for BDHTT-BBT molecule, which suggests the Langevin transport in this molecular system.

Importantly, using ideality factor ($N_{id}$) of particular molecular systems decides the typical transport, either it follows the Langevin type ($N_{id} \rightarrow 1$; trap-free diffusion current) or Shockley–Read–Hall (SRH) mechanism ($N_{id} \rightarrow 2$; trap-assisted recombination).[32,33] The trap-free diffusion mechanism is usually involved in charge transporting devices (e.g., FET and PVs); on the other hand, the light emitting devices (e.g., LEDs) follow the mechanism of SRH recombination. The interesting observation here is that the DOS proportion principally influences the mobility, conductivity, current density and other extended transport quantities. By controlling the DOS proportion, the device performance can be tuned via appropriate parameters, according to the application requirement.

### 4. Entropy-Ruled Boltzmann Approach: Role of DOS Proportion

The DOS proportion $\left(\frac{dh_{eff}}{d\eta}\right)$ principally connects quantum and classical transport regime for wide thermodynamic situations, which is the equivalent form of "*inverse of quantum-classical transition analogy* (QCTA)".[32] The QCTA is nothing but the entropy scaled chemical potential. According to this approach, the change in energy per unit entropy describes electronic property for a given system. If the change in energy (or chemical potential) with respect to entropy is in the order of Fermi energy, the system will be classified as quantum transport system. On the other hand, if the change energy (or chemical potential) of the system over a unit entropy interval is in the order of $k_BT$, that particular system categorizes as the classical transport system. The mixed quantum-classical property of a system is generalized by the chemical potential variation per unit entropy. According to the microscopic structure of any electronic system, one-to-one variation between the chemical potential and entropy will happen, which directly helps to classify the quantum and classical transport transition for any Fermionic systems (from molecules to materials).[32,35] Here, the thermal disorder and bias-driven degeneracy effects on charge carrier dynamics in any Fermionic systems are quantified through thermodynamic entropy ($S$) and differential entropy ($h_S$), respectively. The temperature-lattice vibration relation along with the thermal polaron scattering have a close association with the magnitude of entropy term $S$. For large thermal vibration (vibronic coupling) than electronic coupling (i.e., $k_BT \gg \eta$), the breakdown of translational symmetry is anticipated, and hence the many local minima in the energy landscape (or potential energy surface, PES) is produced, which causes the barrier for electron transport. Depending upon the magnitude of energetic disorder (weak or strong), the charge localization property has been characterized. Here, the energy difference between the adjacent site energies of PES can be accounted for by the chemical potential. That is, the chemical potential is defined as a minimum required energy to activate for charge hopping over the barrier height.[32] In such disordered cases, the carrier transport can be achieved by the thermally activated hopping mechanism, at which $S$ will be a dominant one, $h_{eff} \approx -S$. At this juncture, $S >> h_S$, the expected diffusion current is limited by thermal disorder, which is in agreement with Troisi's approach and is more suitable for molecular systems at high temperature regions of $T \geq 150K$. On the other hand, differential entropy $h_S = \ln(\sqrt{2\pi e}\sigma_{GW})$ has a direct connection with the electronic structure (quantum picture) of molecular or material systems. As discussed in the earlier section, the magnitude of $\sigma_{GW}$ depends on the carrier distribution in the existing possible electronic states within the systems, which is the deciding factor for typical transport; whether it follows localization or delocalization or intermediate transport. We anticipate maximum $\sigma_{GW}$ while the system has more populated energy levels (or DOS), leading towards the delocalized-kind transport. The $\sigma_{GW}$ can be tuned by the applied bias, the bias-driven degeneracy is crucial for enhancing $h_S$, which can help to categorize the field-response transporting materials (e.g., FET).[32,35] For instance, the chemical potential-driven differential entropy is plotted in Fig. 4, which gives a DOS proportion and hence µ/D value. For large $h_S$ ($h_S >> S$), the existing electronic levels are closely continuum (band) and electron/hole dynamics highly correlated with the electronic contribution to coupling (orbital overlapping). The cooperative effect by thermal and electronic counterparts in any electronic systems are fundamentally described by the one-to-one variation between $h_{eff}$ and $\eta$ (or via DOS proportion). In other words, the existence of effective entropy $(h_{eff} = h_S - S/k_B)$ per unit chemical potential is a principal descriptor for all transport quantities such as, DOS, mobility, conductivity, etc. It is important to note that for ideal quantum systems, the chemical potential will be in the order of Fermi energy ($E_F$). At the same time, the chemical potential for classical systems is dominantly weighed by the thermal energy, $k_B T$. Therefore, the DOS proportion for quantum and classical systems are described as $\frac{dh_{eff}}{d\eta} \to \frac{dh_S}{dE_F}$ and $\frac{dh_{eff}}{d\eta} \to \frac{dS}{k_B dT}$, respectively.

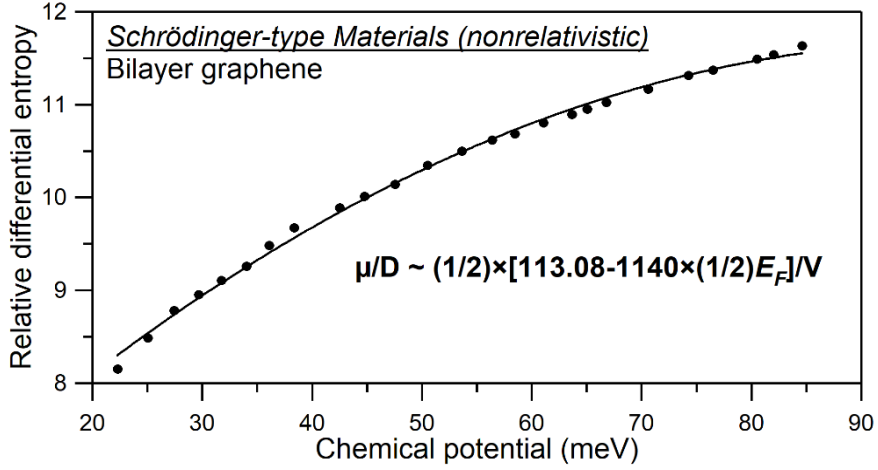

(a)

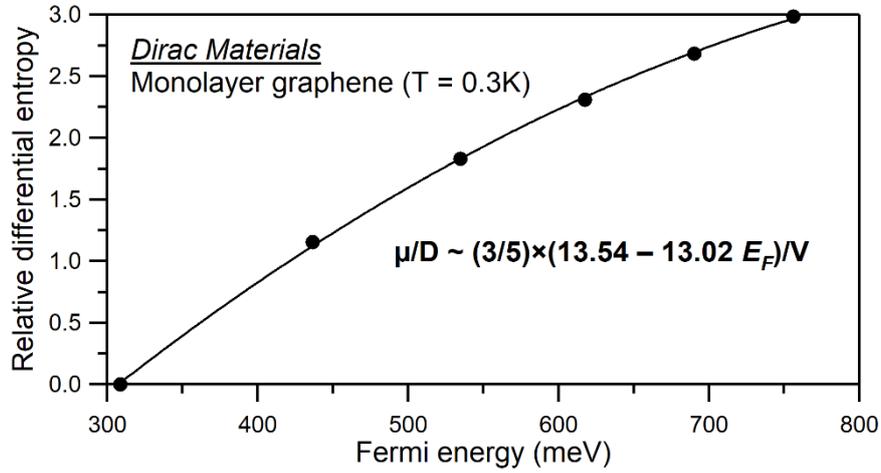

(b)

**Figure 4:** Fermi energy (or chemical potential)-driven differential entropy for mano and bilayer graphene, which results the diffusion-based mobility through the DOS proportion, $\frac{dh_{eff}}{d\eta}$. The detailed study was reported in ref. 35.

The Boltzmann treatment on charge mobility in the periodic systems is expressed as,[32,41]

$$\mu = \frac{qv_F^2 \tau_{rel}}{2n} DOS \qquad (12)$$

According to the entropy-ruled method, the equivalent form of DOS for different dimensional systems is $\left(\frac{d}{d+2}\right)\left(\frac{dh_{eff}}{d\eta}\right)n$, and hence the above mobility relation is redefined as,[35]

$$\mu = q\left(\frac{d}{d+2}\right)\left(\frac{dh_{eff}}{d\eta}\right)\frac{v_F^2 \tau_{rel}}{2} = q\left(\frac{d}{d+2}\right)\left(\frac{dh_{eff}}{d\eta}\right)\frac{l^2}{2\tau_{rel}} \qquad (13)$$

where, $v_F$, $\tau_{rel}$ and $l$ are the Fermi velocity, relaxation time and mean free path, respectively. In principle, the diffusion coefficient for ordered material systems is equated by $D = \frac{v_F^2 \tau_{rel}}{2} = \frac{l^2}{2\tau_{rel}}$. In degenerate quantum materials, the electronic levels are continuum and hence the expected charge transfer kinetics is in the order of Fermi energy. That is, within the Fermi surface region, the carrier energy assumed as $E_F = \frac{\hbar^2 k_F^2}{2m^*} \to \frac{1}{2} m^* v_F^2$. Within this infinitesimal energy gap, the charge drift dynamics or flux can be associated with the relative chemical potential along the wrapped Fermi surface of the given materials. Here, the chemical potential is the total energy per particle, which gives an average energy of a system. The general form of average energy ($E_{av}$)-Fermi energy ($E_F$) relationship as given by $E_{av} = \left(\frac{d}{d+2}\right) E_F$. In such that, the Fermi energy-electron density relation for 1D, 2D and 3D degenerate periodic systems are $E_F = \frac{\pi^2}{4m^*} n_{1D}^2$, $E_F = \frac{\pi \hbar^2}{2m^*} n_{2D}$ and $E_F = \frac{\hbar^2}{2m^*} \left(3\pi^2 n_{3D}\right)^{2/3}$, respectively. Besides that, for a delocalized band transport model, $dh_S = \frac{d\sigma_{GW}}{\sigma_{GW}} \approx \frac{\sigma_{GW,f} - \sigma_{GW,i}}{\sigma_{GW,f}} \to 1$; since $h_S = ln(\sqrt{2\pi e}\sigma_{GW})$. With these assumptions, the above entropy-ruled Boltzmann mobility is now reduced as (see refs. 32, 35),

$$\mu = \frac{q\tau_{rel}}{m^*} \qquad (14)$$

This is the conventional mobility equation for band transport systems. Here, the nonequilibrium and degeneracy effects due to applied bias can be accounted for by differential entropy. In this extent, one can also arrive the conventional mobility equation from the entropy-ruled method for Dirac systems, $\mu_{Dirac} = \frac{qv_F^2 \tau_{rel}}{E_F} \equiv \frac{qv_F \tau_{rel}}{\hbar k_F} \to \frac{ql}{\hbar k_F}$; since $E_F = \hbar v_F k_F$. The more accurate mobility calculation can be achieved by Eqn. 13, which is a unique formalism for both relativistic particle dynamical systems (Dirac materials, $E = \hbar v_F k$), as well as nonrelativistic particle dynamical systems (solved by Schrödinger equation, here $E = \frac{\hbar^2 k^2}{2m}$). For periodic materials, the diffusion coefficient generally has direct proportional relationship with the square of mean free path ($l$), $D = \frac{v_F^2 \tau}{2} \equiv \frac{l^2}{2\tau}$. On the other hand, the diffusion coefficient for disordered molecular or material systems can be quantified by $D = \frac{1}{2d} \frac{\Delta \langle R^2(t) \rangle}{\Delta t}$, where $\langle R^2(t) \rangle$ is the mean squared displacement and $t$ is the simulation time.

## 5. Validity and Limitations of Einstein's Relation: An Entropy-Ruled Approach

Einstein's diffusion-mobility (D/μ) is a basic transport relation to explore the mobility in disordered systems (e.g., molecules). In such energetic disordered cases, the CT follows the thermally activated hopping mechanism, in which the diffusion constant is obtained from the mean squared displacement over the time. Here, the diffusion process along the random site energy landscape is simulated using a random walk procedure. At high temperature ($T \geq 150K$), the large amplitude of thermal vibration (nuclear dynamics) breakdowns the translational symmetry of molecular geometry, which causes many local minima in the PES.[8] Here, each local minima acts as a one energy site and hence the collection of many site energies in PES is expected that reveals the energy barrier for the charge hopping process. In this case, the Einstein's D/μ relation has been generally used for charge transport measurement and calculations, which has a linear relationship with the temperature, $\frac{D}{\mu} = \frac{k_B T}{q}$. Over the last some decades, several experimental and theoretical studies clearly indicate that the failure of classical Einstein diffusion-mobility relation is due to nonlinear transport phenomena (see refs. 22, 35, 36; and cross references therein). The Einstein relation does not give more accuracy for high charge density degenerate systems/devices and also for nonequilibrium cases under strong applied electric field conditions. From the past five decades, the possibility of theoretical as well as experimental attempts are made to explain the quantum feature and drift-coupled diffusion-based mobility calculation in various systems, but still not well parameterized. To drag the electronic structure information, it is needed to probe quantum dynamics and quantify the nonlinear phenomena due to environmental (or external) interactions in degenerate molecular and material semiconductors. In this scenario, the entropy-ruled charge transport method has been proposed to overcome these disadvantages of the classical Einstein relation of D/μ. By implementing the entropically varying charge density formalism into the generalized Einstein relation, one can attain the entropy-ruled D/μ relation as described by,[35]

$$\left(\frac{D}{\mu}\right)_d = \left[\left(\frac{d+2}{d}\right)\frac{1}{q}\frac{d\eta}{dh_{eff}}\right] \qquad (15)$$

This is the generalized version of Einstein relation for both band and hopping transport systems and it works the entire range from equilibrium to nonequilibrium. For molecular transport, the chemical potential is arrived from the charge transfer reaction mechanism in the molecular systems/species, and hence the activity (*a*)-chemical potential (*η*) relation is generally written as,[32,35,42]

$$\eta = k_B T \ln a + \eta_0 \equiv k_B T (\ln a + 1) \qquad (16)$$

At high temperature, the molecular transport is explained by the thermally activated hopping mechanism and hence the chemical potential is $\eta_0 \equiv k_B T$. Here, the chemical activity is the activity coefficient ($\zeta$) times of concentration ratio (or particle density ratio), i.e., $a = \frac{n}{n_0}\zeta$. In this case, the activity coefficient is $\zeta = exp\left[\frac{(\Delta E_{ij}+\lambda)^2}{4\lambda k_B T}\right]$, which is a key factor for electron hopping process; where, $\Delta E_{ij}$, $\lambda$ are the site energy difference between adjacent sites and reorganization energy, respectively. In this charge transfer reaction, the minimum required energy to hop across the barrier height between the adjacent potential energy surfaces must be equal to $\frac{(\Delta E_{ij}+\lambda)^2}{4\lambda} = \frac{\lambda}{4}\left(1 + \frac{\Delta E_{ij}}{\lambda}\right)^2$. Now, the chemical potential is $\eta = k_B T(\ln a + 1) = k_B T\left[\ln\left(\zeta \frac{n}{n_0}\right) + 1\right]$. By inserting the above relations, along with Eqn. 6 into Eqn. 16, the generalized form of the chemical potential for different dimensional ($d$ = 1D, 2D and 3D) molecular systems can be expressed as

$$\eta_d = k_B T \left(1 + \left(\frac{d}{d+2}\right)h_{eff} + \frac{(\lambda+\Delta E(\vec{E}))^2}{4\lambda k_B T}\right) \qquad (17)$$

Using the chemical potential, the governed equation of entropy-ruled diffusion coefficient-mobility relation as,

$$\left.\frac{D}{\mu}\right|_d = \left(\frac{d+2}{d}\right)\frac{k_B T}{q}\left[\left(\frac{d}{d+2}\right) + \frac{(\lambda+\Delta E(\vec{E}))}{2\lambda k_B T}\frac{dE(\vec{E})}{dh_{eff}}\right] \qquad (18)$$

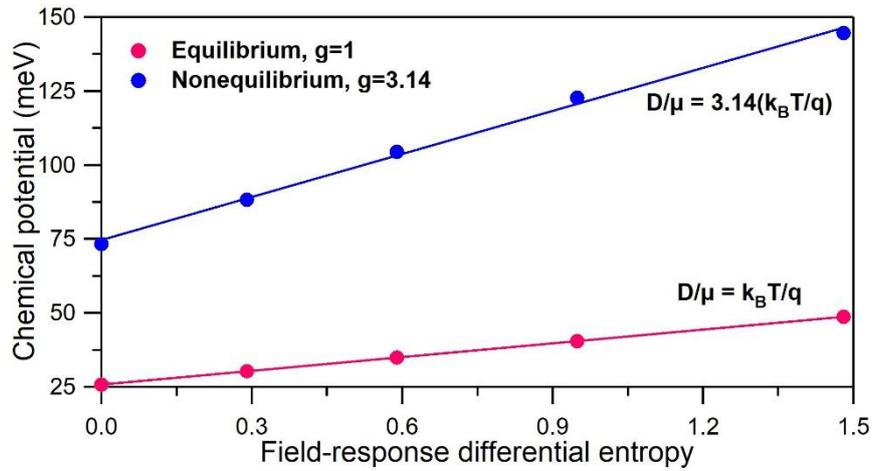

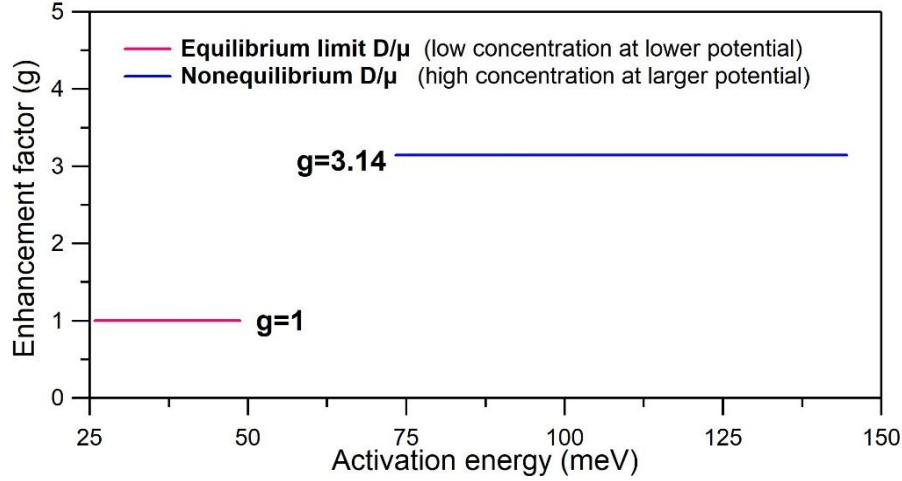

**Figure 5:** The validity and limitations are studied through BDHTT-BBT molecular solids at different bias (electric field) condition. The deviation of Einstein's relation is equated by enhancement factor (*g*). The results show that for the large carrier' energy flux and degenerate systems, the enhancement has been observed. The calculation procedures were given in ref. 32.

In this relation, the term $\frac{dE(\vec{E})}{dh_{eff}}$ is nothing but the entropy scaled energy. The thermodynamic coupled degeneracy weightage is related by the effective entropy. For molecular/organic devices, the electric field assisted carrier energy flux along the consequential hopping sites can be accounted for by $\frac{dE(\vec{E})}{dh_{eff}}$, which is also responsible for nonequilibrium transport in which Einstein's relation takes the deviation from its original value, $k_BT/q$. This deviation value is compensated by the enhancement factor (*g*), i.e., $\frac{D}{\mu} = g\frac{k_BT}{q}$. The *g* value is always greater than unity for nonequilibrium and degenerate transport systems (see Fig. 5). The generalized form of entropy-ruled relation (Eqn. 18) will be reduced as an original Einstein's relation $\left(\frac{D}{\mu} = \frac{k_BT}{q}\right)$, when $\frac{dE(\vec{E})}{dh_{eff}} \to 0$, which is an equilibrium transport relation for disordered systems with high temperature range which is shown in Fig. 5. In such a case, the *g* value will always be unity. The main point here is that the conventional Einstein's classical relation does not depend on the dimension of the system (1D, 2D and 3D), i.e., same expression for the whole system $D/\mu = k_BT/q$. On the other hand, our proposed entropy-ruled $D/\mu$ relation has dimension-dependent, specifically dimension-dependent density of states (DOS) play a crucial role for this relation.

## 6. Future Scope and Conclusion

The entropy-ruled conceptualization on charge carrier dynamics from molecule to materials provides a new perception on fundamental transport mechanism in different solid-state systems. Importantly, it unifies band and hopping transport mechanism and it works from low to high temperature ranges in both equilibrium to nonequilibrium cases. The interesting point is that a newly introduced DOS proportion is principally acting as a key descriptor for all basic CT quantities such as, mobility, conductivity, current density, etc. Also, it is further expected that the entropy scaled chemical potential helps to explore some interesting new strange behaviour in different Fermionic systems, which might give a fresh look on charge dynamics in both interacting and noninteracting systems. In this regard, the entropy-ruled method along with the quantum phase transition will be a crucial concept, which can act as a basic describing principle for next level understanding in electron dynamics, and to further explore a new functional-driven potential applications (e.g., electronic devices, charge storage devices). Notably, for certain combinations of temperature and chemical potential, we reproduce the original Einstein equation from the entropy-ruled method. Through this method, one can elucidate the electronic contribution on thermoelectricity in different quantum materials, which explains the validity and limitations of Wiedemann-Franz law and Mott relation. Noteworthy, this study is very important to get a clear understanding of carrier dynamics in a different range of systems of the device interest and related process, from molecules to materials, at different physical domains. However, the present entropy-ruled method does not include the exchange-correlation effect on charge transport in strongly correlated systems, which is important for spin (also collective behaviour) transport and many-body quantum D/µ transport relation and related device applications. Therefore, future studies need to give more attention to the DOS proportion from the perspective of quantum many-body localization (including spin effect). Using Eqn. 3, the total interaction effects (including external influences such as voltage, temperature and doping) can be well-approximated by the effective entropy ($h_{eff} = h_S - S/k_B$). The entropy correlated energy difference (between adjacent nearby electronic levels) is good-approximated here by $\Delta E(h_{eff}) \approx \Delta E_{initial} \exp\left(-\frac{h_{eff}}{2}\right)$.[35] This additional correction energy will be included in the total energy ($E^* = E + \Delta E$), which influences the carrier density distribution too. Therefore, the expected D/µ relation for strong interacting electronic systems can be approximated by $\left(\frac{D}{\mu}\right)_d = \frac{1}{\alpha_d}\left(\frac{d+2}{d}\right)\frac{1}{q}\frac{\Delta \eta}{\Delta h_{eff}}$; where, $d$ and $\alpha_d$ are the dimension (1D, 2D and 3D) and fitting parameter, which is the dimension-dependent (1D, 2D and 3D) parameter. At $\alpha_d \rightarrow 1$, one can get the D/µ relation for free electron systems. In this extent, we can introduce the imperfect Fermi-Dirac (IFD) distribution function (using entropy correlated energy gap equation) as follows,[35]

For electron, $$f(E,T,h_{eff}) \approx \frac{1}{1+\exp\left[\beta(E-\eta)\exp\left(-\frac{h_{eff,e^-}}{2}\right)\right]};  \quad (19.a)$$

For hole, $$f(E,T,h_{eff}) \approx \frac{1}{1+\exp\left[\beta(\eta-E)\exp\left(-\frac{h_{eff,h^+}}{2}\right)\right]}  \quad (19.b)$$

Here, $\beta$ and $\eta$ are the Lagrange multiplier ($1/k_BT$) and chemical potential, respectively. For degenerate systems, the energy gap (here, $\Delta E=E-\eta$) between the electronic levels is minimum.[43] The electronic coupling (orbital level interactions) increases while the energy gap between the electronic states decreases and vice-versa. Here, the differential entropy ($h_S$) favors the electronic coupling; on the other hand, the thermal disorder (in terms of thermodynamic entropy, $S$) minimizes the electronic coupling, but it increases the vibronic effect. In this context, one can say that the energy difference or gap is reduced by $h_S$ and increased by $S$ (due to large thermal vibration). Using this entropy-committed approximation (for energy variation (or energy difference)), we can extend the charge transport calculation from the free-electron model to strongly interacting Fermi systems; accordingly, carrier effective mass and deformation potential will be modified. This variation directly influences the mobility and other transport quantities. For highly degeneracy and correlated electron transport systems, the IFD distribution function will be playing a vital role to calculate the charge density, electronic compressibility (or DOS), energy-wavevector relation and effective mass. Through the entropy-ruled transport approach, the theoretical evaluation of electron temperature in particular molecules or material is fundamentally important and solicited to model the best electron dynamical systems, which might be a great interest for developing a new era in quantum features' electron devices. Henceforth, the systematic development of entropy-ruled charge and energy transport method have to be further essential to probe the exact electron dynamical picture under different physical situations.

**Notes**

The author declares no competing financial interest.